\documentclass[fleqn,usenatbib,useAMS]{mnras}

\usepackage{graphicx}	
\usepackage{amsmath}	
\usepackage{amssymb}	
\usepackage{multicol}        
\usepackage{bm}		
\usepackage{pdflscape}	
\usepackage{comment}
\usepackage{xcolor}

\usepackage{svg}
\newcommand{\orcid}[1]{\href{https://orcid.org/#1}{\includesvg[width=10pt]{orcid}}}

\usepackage{multirow}

\usepackage[T1]{fontenc}
\usepackage{ae,aecompl}

\usepackage{newtxtext,newtxmath}

\title[Identifying BLAPs from Gaia \& ZTF]{Identifying Blue Large Amplitude Pulsators from Gaia DR2 \& ZTF DR3}

\author[P. R. McWhirter et al.]{Paul Ross McWhirter$^{1,2,3\,\orcid{0000-0002-6616-0782}}$ and
Marco C. Lam$^{4\,\orcid{0000-0002-9347-2298}}$\thanks{Contact e-mail: \href{mailto:lam@tau.ac.il}{lam@tau.ac.il}}\\
$^1$Astrophysics Research Institute, Liverpool John Moores University, IC2, LSP, 146 Brownlow Hill, Liverpool L3 5RF, U.K.\\
$^2$Instituto de Astrof\'isica de Canarias~(IAC), Calle V\'ia L\'actea s/n, E-38200 La Laguna, Tenerife, Spain\\
$^3$Universidad de La Laguna, Dept. Astrof{\'i}sica, E-38206 La Laguna, Tenerife, Spain\\
$^4$School of Physics and Astronomy, Tel Aviv University, Tel Aviv 69978, Israel}

\date{Last updated \today}

\pubyear{2021}

\begin{document}
\label{firstpage}
\pagerange{\pageref{firstpage}--\pageref{lastpage}}
\maketitle

\begin{abstract}
Blue Large Amplitude Pulsators~(BLAPs) are hot, subluminous stars undergoing rapid variability
with periods of under $60$\,mins. They have been linked with the early stages
of pre-white dwarfs and hot subdwarfs. They are a rare class of variable star due to
their evolutionary history within interacting binary systems and the
short timescales relative to their lifetime in which they
are pulsationally unstable. All currently known BLAPs are relatively faint
($15-19$\,mag) and are located in the Galactic plane. 
These stars have intrinsically blue colours but the large interstellar extinction
in the Galactic plane prevents them from swift identification using colour-based
selection criteria. In this paper, we correct the Gaia $G$-band apparent magnitude
and $G_{\mathrm{BP}}-G_{\mathrm{RP}}$ colours
of $89.6$ million sources brighter than $19$\,mag in the Galactic plane with
good quality photometry combined with supplementary
all-sky data totalling $162.3$ million sources.
Selecting sources with colours consistent with the known population of BLAPs and performing a
cross-match with the Zwicky Transient Facility~(ZTF) DR3, we identify
$98$ short period candidate variables. Manual inspection of the period-folded light curves
reveals $22$ candidate BLAPs. Of these targets, $6$ are consistent with the observed periods and
light curves of the known BLAPs, $10$ are within the theoretical period range of
BLAPs and $6$ are candidate high-gravity BLAPs. We present follow-up spectra of $21$ of these
candidate sources and propose to classify $1$ of them as a BLAP, and tentatively assign 
an additional $8$ of them as BLAPs for future population studies.

\end{abstract}




\begingroup
\let\clearpage\relax
\endgroup
\newpage

\section{Introduction}
Blue Large Amplitude Pulsators (BLAPs) are hot, subluminous variable stars which was
first identified in the Optical Gravitational Lensing Experiment IV~\citep[OGLE-IV, ][]{2015AcA....65....1U}), and subsequently
the high-gravity (high-g) BLAPs with the Zwicky Transient Factory~(ZTF) survey data~\citep[hereafter P17,
K19]{2017NatAs...1E.166P, 2019ApJ...878L..35K}. P17 presented the first $14$ BLAPs,
which were all detected in OGLE fields towards the Galactic bulge. One of which was previously
classified as a peculiar $\delta$~Scuti star~\citep{2017MNRAS.465..434M}, which is a type of
main sequence A-type stars that pulsate ~\citep{1900ApJ....12..254C}.
The BLAPs are of spectral type B
and exhibit periods shorter than those of $\delta$~Scuti variables at~($20-40$\,mins)
and higher amplitudes for these
periods: $0.19-0.36$\,mag in the $I$-band and $0.22-0.43$\,mag in the $V$-band.
Their light curves resemble those of fundamental-mode Cepheid
variables with asymmetry in the brightening and dimming phases.
They brighten rapidly from minimum to maximum brightness over a
duration of $\approx25\%$ of their period
before a slower decline which sometimes exhibits a `step' where
the rate of fading abruptly decreases.
Their high amplitudes and saw-tooth light curve shape suggest they are
pulsating in a low-order radial mode although it is not conclusive. Some BLAP models have been found to
be pulsationally unstable in high-order non-radial modes
although with a weaker excitation than that of the fundamental mode~\citep{2018arXiv180907451C}.
K19 identified $4$ additional variable stars which differed from the previously discovered BLAPs
by exhibiting shorter periods ($3-8$\,mins) and lower amplitudes ($0.05-0.1$\,mag in
the $r_{\mathrm{ZTF}}$ band) classifying them as high-g BLAPs. The OGLE survey also included multiple fields
aimed at the Magellanic Clouds and follow-up study of these fields has revealed no
BLAPs~\citep{2018pas6.conf..258P}, although the distance moduli to these galaxies would make
their detection difficult. Thus, they could only have ruled out the lack of foreground BLAPs
in our Galaxy. High cadence photometry of OGLE-BLAP-009 and OGLE-BLAP-014 have not revealed any
additional periodicity down to periods of $20$\,s, indicating BLAPs likely only have a
single radial pulsation mode~\citep{2020MNRAS.496.1105M}. Non-radial gravity modes have not yet been observed in BLAPs. The OGLE BLAP light curves are of sufficient baseline to detect higher-amplitude gravity modes as they would have a longer period than the fundamental or radial overtone modes assuming the visible pulsation is the fundamental mode.

The spectroscopic followup by P17 and K19 confirmed that BLAPs are substantially hotter
than $\delta$~Scuti variables with effective temperatures of
$\mathrm{T}_{\mathrm{eff}}\approx20\,000-35\,000$\,K, surface gravity of 
$\log(g / {\rm cm\,s^{-2}}) = 4.2 - 4.6$ for the OGLE BLAPs and 
$\log(g / {\rm cm\,s^{-2}}) = 5.3 - 5.7$ for the high-g BLAPs.
The OGLE BLAPs also exhibit moderate helium enrichment. 
Their spectra are of early B-type/late O-type stars,
similar to the hot subdwarfs but with surface gravity ten times lower and higher luminosity,
indicating that they are in a giant configuration.
The high-g BLAPs are more compact and have luminosity and surface gravity consistent with hot subdwarfs.
P17 and K19 show there is a clear periodic colour change as a function of the oscillation phase 
epoch-folded at the dominant period suggesting
pulsation is the cause of the variability.
The longer period BLAP light curves demonstrate the characteristic high-amplitude sawtooth in
the $g$ and $r$ bands seen in radially pulsating stars~\citep{1991PASP..103..933M}.

The BLAP variability is theorised to be a result of pulsations due to the $\kappa$-mechanism driven
by the partial ionisation of iron-group metals in a shell of elevated abundance due to 
radiative-levitation~\citep{2018MNRAS.481.3810B}.
These pulsations occur when stars of given composition and
structure expand and contract their outer layers radially to maintain
equilibrium~\citep{1917Obs....40..290E}; or due to non-radial variations in surface
temperature~\citep{1989nos..book.....U}. They can be used to probe the inner structure of
these variable stars~\citep{1984ARA&A..22..593D}. BLAPs have been linked with the early
stages of $0.27-0.37\mathcal{M}_{\sun}$ Extremely Low Mass (ELM)
pre-White Dwarfs (pre-WDs) and $0.45\mathcal{M}_{\sun}$ hot subdwarfs which
both evolve through the instability region associated with this type of
pulsation~\citep{2018MNRAS.477L..30R, 2018MNRAS.481.3810B, 2018MNRAS.478.3871W, 2019ApJS..243...10P}.
They have also been proposed to be the surviving remnants of single-degenerate Type Ia
supernovae~\citep{2020ApJ...903..100M}.
Mixing due to some combination of convection, rotation and shell-flashes can result in the
stellar surface containing a mixture of hydrogen and helium explaining the helium surface
abundance of the OGLE BLAPs. High-g BLAPs are more compact, and therefore gravitational settling
has removed this excess helium from the photosphere.
\citet{2020MNRAS.492..232B} produced \textsc{MESA} stellar models for
pre-WDs with masses from
$0.18\mathcal{M}_{\sun}$ to $0.46\mathcal{M}_{\sun}$ with the effects of radiative levitation.
They identified an extended region of iron-group instability using a non-adiabatic analysis of the 
\textsc{MESA} models at high time resolution (see Section 2 of \citealt{2020MNRAS.492..232B} for details).
The observed periods of the BLAPs from P17 and the high-gravity BLAPs are
consistent with the fundamental modes of \textsc{MESA} pre-WD models.
The oscillation analysis indicates that pre-WDs with
effective temperatures of up to $\mathrm{T}_{\mathrm{eff}}\approx50\,000$\,K can exhibit pulsations.
Some of these models also show evidence of instability in higher-order pressure modes which would
exhibit shorter periods than the fundamental mode.

BLAPs appear to be a rare class of variable star due to their evolutionary history in an
interacting binary system and the short timescales in which they transition the instability
strip~\citep{2020MNRAS.492..232B}.
Binary Population Synthesis modelling has demonstrated that BLAPs form within interacting binary
systems and can be both the primary with a main sequence companion or a secondary
with a compact companion~\citep[hereafter BSE21]{2021MNRAS.507..621B}.
Within the BSE21 simulated population, a small percentage of binaries also
evolved to a type Ia supernova resulting in an isolated BLAP as previously predicted.
The remaining BLAPs within the BSE21 population transition the instability strip with
a companion with a mass often below $1.5\mathcal{M}_{\sun}$ with a majority of companions
having masses of $0.4-0.6\mathcal{M}_{\sun}$. 
Roche-Lobe Overflow is the dominant mechanism of mass transfer resulting in
wider orbits with periods of $>1$\,day.
The population of BLAPs within the analysis of BSE21 also evolved to form two distinct populations
as seen in the OGLE BLAPs and the high-g BLAPs suggesting that this characteristic of these
variable stars is not a result of observational bias.
BSE21 considers this synthetic BLAP population within a number of models of the Milky Way disk and
conclude the number of BLAPs visible in the Milky Way at the present time as $\approx12\,000$.

There is a large discrepancy between the number of BLAPs observed in current surveys and the number
predicted by synthetic BLAP populations in Galactic models after considering the magnitude limits
of these surveys.
All currently known BLAPs are relatively faint
($15-19$\,mag) and are located in the Galactic plane. These stars
have intrinsically blue colours but the interstellar extinction due to dust in the Galactic plane
prevents these sources from swift identification using colour-based selection criteria. 
\citet{2018A&A...620L...9R} demonstrates a method of correcting this interstellar extinction
for the BLAPs in P17 using
all-sky dust maps combined with Gaia DR2 astrometry. 

We structure the paper as follows. In \textsection2 we define the criteria used in the selection
of our initial Gaia DR2 sources. We then detail the method used to
correct the interstellar extinction of these sources and detect periodic variability
to reveal the population of intrinsically blue, subluminous variable stars.
These candidate variable stars are studied for BLAP-like properties and we
present these new BLAP candidates.
In \textsection3 we describe a follow-up spectroscopic campaign on these $22$ BLAP candidates and
identify features consistent with spectra from the known BLAP population.
Finally, in \textsection4 we discuss and conclude the results of this work and propose further
spectroscopic campaigns to reveal the stellar properties of these stars for future
theoretical analysis. In this paper, the term \textit{amplitude} refers to the minimum to maximum
variation in magnitude unless otherwise stated.

\section{Data Selection Criteria}

Gaia DR2~\citep[GDR2; ][]{2018A&A...616A...1G} is a 1.7-billion-source catalogue with an
unprecedented astrometric quality accompanied with photometry in three filters, $G$,
$G_{\mathrm{BP}}$ and $G_{\mathrm{RP}}$. The accuracy in the bright end~($<14$\,mag)
is as high as 0.02\,mas in the parallax, while in the faint end ($\sim$ $21$\, mag), the
precision is about $100$ times larger. The high-level pre-processed data can be retrieved
via their dedicated archive page\footnote{\url{https://gea.esac.esa.int/archive/}}.
\citet{2018AJ....156...58B} computed the distance of 1.33 billion stars from their parallaxes
using a weak distance prior that varies smoothly as a function of Galactic longitude and
latitude according to a Galaxy model.

Pan-STARRS 1~(PS1)\footnote{\url{https://panstarrs.stsci.edu/}} is a $1.8$\,m wide-field optical
imager on the peak of Haleakala on Maui~\citep{2016arXiv161205560C}. The 3$\pi$ Survey covers
$30\,000\,$deg$^2$ of the sky north of $-30^{\circ}$ in declination~\citep{2020ApJS..251....6M}.
It imaged thse sky on average $60$ times in the $g$, $r$, $i$, $z$, and $y$-band filters over the
$3.5$ year survey period. One affiliated product from this enormous ground-based survey is the 3D
dust map~\citep{2019ApJ...887...93G}\footnote{\url{http://argonaut.skymaps.info/}}.
The map was produced using $800$ million stars from PS1 and 2MASS~\citep{1997ASSL..210...25S}, and
included the Gaia parallaxes to improve the distance estimates to nearby stars for more accurate
reddening estimation. They also incorporated a spatial prior that correlates the dust density across
nearby sightlines to produces a smooth dust map. Their reported reddening uncertainties are typically
$\sim$ $30\%$ smaller than those from the GDR2.

Zwicky Transient Facility~\citep[ZTF, ][]{2019PASP..131a8002B} is an optical time-domain survey
that uses the Palomar 48-inch Schmidt telescope. Each exposure cover $47$\,deg$^{2}$ in one of
the $g$, $r$ or $i$-band filters. In the DR3, $\sim$ $1.4$ billion sources with more than
$20$ detections in the northern sky with a survey footprint similar to that of the PS1 3 $\uppi$
Survey, have become publicly available\footnote{\url{https://irsa.ipac.caltech.edu/Missions/ztf.html}}
with
new data release scheduled every six months.

Our data selection starts with queries on the GDR2 archive. To define the quality requirements
for our initial dataset we utilise the treatise defined for astrometric quality in
\citet{2018A&A...616A...2L} and photometric quality in \citet{2018A&A...616A...4E} with queries
based on those from GDR2. Due to the declination limit of the PS1 3D dustmaps our footprint is limited
to targets with declination $>-30^{\circ}$. We further constrain this declination to $>-15^{\circ}$ to
ensure optimal placement for observation from northern facilities. GDR2 focused on the production of a
GDR2 Hertzsprung-Russell diagram of nearby stars and require the astrometric signal-to-noise ratio to
be larger than 10. Our selection relaxes this requirement to be greater than $1$ as the BLAPs are
expected to be faint and much more distant. We also limit our sample to those with parallaxes less
than or equal to $1$\,mas\,yr$^{-1}$ to remove nearby hot WDs which have a fainter absolute magnitude. For
the photometric quality, we consider the faint nature of our targets. We limit our sample to sources
brighter than $19$\,mag and apply relaxed constraints relative to GDR2 requiring the mean flux
significance to be $G/\sigma_{G}>5$ and the mean flux significance of  $G_{\mathrm{BP}}/\sigma_{G_{\mathrm{BP}}} > 20$ and
$G_{\mathrm{RP}}/\sigma_{G_{\mathrm{BP}}} > 20$. The final quality selection criteria are identical to those in GDR2 as per recommendations from~\citet{2018A&A...616A...4E} related to
blending between sources or the background in the $G_{\mathrm{BP}}$ and $G_{\mathrm{RP}}$ bands. This
blending has been used to calculate a factor named the $G_{\mathrm{BP}} - G_{\mathrm{RP}}$ excess
factor of which fainter sources are more strongly affected. These criteria are an empirically
determined non-linear function of the $G_{\mathrm{BP}} - G_{\mathrm{RP}}$ colours
of the sources:
\verb+phot_bp_rp_excess_factor+
$> 1.0 + 0.015 (G_{\mathrm{BP}} - G_{\mathrm{RP}})^2$ and 
\verb+phot_bp_rp_excess_factor+
$< 1.3 + 0.06 (G_{\mathrm{BP}} - G_{\mathrm{RP}})^2$.

The above conditions are applied to the GDR2 archive using Astronomical Data Query
Language~(ADQL) queries. We apply one final condition to the query for targets
within $-15^{\circ} < b < 15^{\circ}$ for sources
in the Galactic plane. We assume that outside of the Galactic plane, the substantially lower
quantity of interstellar dust would result in BLAPs retaining their intrinsically blue colours
and therefore being detected with existing surveys for hot, subluminous sources such as those
conducted using GDR2~\citep{2019A&A...621A..38G, 2020A&A...635A.193G}.
The resulting dataset consists of $89\,633\,653$ sources within
the Galactic plane.

After the initial selection of sources, we extended the method to
sources outside the Galactic plane and to all-sky sources with parallax-over-error
values below 1.0 values to produce two supplementary datasets which are
also processed by our method.
The dataset of sources outside of the selection of $-15^{\circ} < b < 15^{\circ}$
contains $35\,498\,294$ sources and the all-sky `low astrometric signal-to-noise'
dataset contains
$37\,245\,589$ sources. This dataset is dominated by faint stars close to the $19$\,mag
limit of the search queries. The combined total of all-sky sources is $162\,377\,536$.

We designate our targets with the unique identification
starting with ZGP, which is comprised
of the first letter of each of ZTF, Gaia and Pan-STARRS. The unique identifier
of the targets is ordered by increasing right ascension.

\subsection{Method}

By combining the accurate parallaxes from the GDR2 and the 3D dust map from the PS1,
we dereddened the $89.6$ million stars brighter than $19$\,mag
within the Galactic plane dataset.
The first operation is to transform the parallaxes of the sources in these datasets to distances
from the Earth within the Galaxy. We utilise the computations from~\citet{2018AJ....156...58B}
described above which are available as a supplementary table within the GDR2 archive.
We do this by using the ADQL queries for the above quality selection and perform a join onto the
\verb+gaiadr2_geometric_distance+ table which contains the required data.
The resulting data contains an estimated distance (\verb+r_est+) and a higher and lower bound
(\verb+r_hi+ and \verb+r_lo+) defined as the distances where the probability of the
targets being within this region is $p=0.6827$, which is within the 1 standard deviation from
the centre of a ``Gaussian distribution''. As the distance posterior is asymmetrical, the difference
between \verb+r_hi+ and \verb+r_lo+ with \verb+r_est+ are unequal.

We take these three distance estimates and compute the interstellar reddening for each distance using
the dustmaps package with the PS1 3D dustmaps~\citep[][hereafter \texttt{Bayestar2019}]{2019ApJ...887...93G}.
This package can be used to compute the reddening locally on the desktop computer or to query an
online resource. We chose to query the online data which is performed three times for each source with
the distance estimates. The reddening provided by \texttt{Bayestar2019} are mapped to PS1 passbands
and are related to $E(g-r)$, the reddening in the PS1 $g$ and $r$ passbands.

To utilise these reddening values on the GDR2 data we must either transform the GDR2 photometry
into the PS1 passbands or convert the extinctions to $E(G_{\mathrm{BP}}-G_{\mathrm{RP}})$. The
first issue with this calculation is a result of the conversion being a function of the intrinsic
colour of a source which is an unknown without spectral information. We apply corrections
computed for the $R_{\mathrm{V}}=3.1$ reddening law from~\citet{1999PASP..111...63F} to
a $7000$\,K source spectrum shown in~\citet{2011ApJ...737..103S}. As a result of using a source
spectrum significantly cooler of those of BLAPs, the reddening will be underestimated for BLAPs.
We mitigate this issue by defining future selection criteria on known BLAPs subjected to this
method. We utilise the first relation in \citet{2011ApJ...737..103S} to compute $E(B-V)$, the
reddening in the Johnson $B$ and $V$ passbands as $E(B-V)=0.884\times$ \texttt{Bayestar2019}.
We utilise $E(B-V)$ for this conversion as existing dustmaps of the Galactic plane have been used
to compute the relation between $E(B-V)$ and $E(G_{\mathrm{BP}}-G_{\mathrm{RP}})$.
\citet{2019MNRAS.483.4277C} computed the relation between $E(B-V)$ and
$E(G_{\mathrm{BP}}-G_{\mathrm{RP}})$ to compare their dustmaps to other research determining that
$E(B-V)$ $=0.75\times$ $E(G_{\mathrm{BP}}-G_{\mathrm{RP}})$. Combining these research findings, we
utilise the computation of $E(G_{\mathrm{BP}}-G_{\mathrm{RP}})$ $=(0.884/0.75)\times$
\texttt{Bayestar2019} to transform the three reddening estimates into the GDR2 system.

With the computation of $A_{G}$ and $E(G_{\mathrm{BP}}-G_{\mathrm{RP}})$, we compute the absolute magnitude, $\mathrm{M}_{G},$ and the dereddened $G_{\mathrm{BP}}-G_{\mathrm{RP}}$ colour, $(G_{\mathrm{BP}}-G_{\mathrm{RP}})_0$
using equations~\ref{eq:mg} and \ref{eq:bprp0}.
\begin{equation}
\mathrm{M}_{G} = \mathrm{m}_{G} - 5 \log{\left(\mathrm{r}_{\mathrm{est}}\right)}-5 - A_{G}
\label{eq:mg}
\end{equation}
where $\mathrm{M}_{G}$ is the absolute magnitude of a source in $G$,
$\mathrm{m}_{G}$ is the apparent magnitude of a source in $G$,
$\mathrm{r}_{\mathrm{est}}$ is the estimated distance to the source in parsecs
and $A_{G}$
is the interstellar extinction of the source in $G$.
\begin{equation}
(G_{\mathrm{BP}}-G_{\mathrm{RP}})_0 = (G_{\mathrm{BP}}-G_{\mathrm{RP}}) - E(G_{\mathrm{BP}}-G_{\mathrm{RP}})
\label{eq:bprp0}
\end{equation}
where $(G_{\mathrm{BP}}-G_{\mathrm{RP}})_0$ is the dereddened $(G_{\mathrm{BP}}-G_{\mathrm{RP}})$ colour of a source, ($G_{\mathrm{BP}}-G_{\mathrm{RP}}$) is the measured
colour of a source from GDR2 and $E(G_{\mathrm{BP}}-G_{\mathrm{RP}})$ is the interstellar reddening of a source.

\begin{figure}
    \centering
    \includegraphics[width=0.5\textwidth]{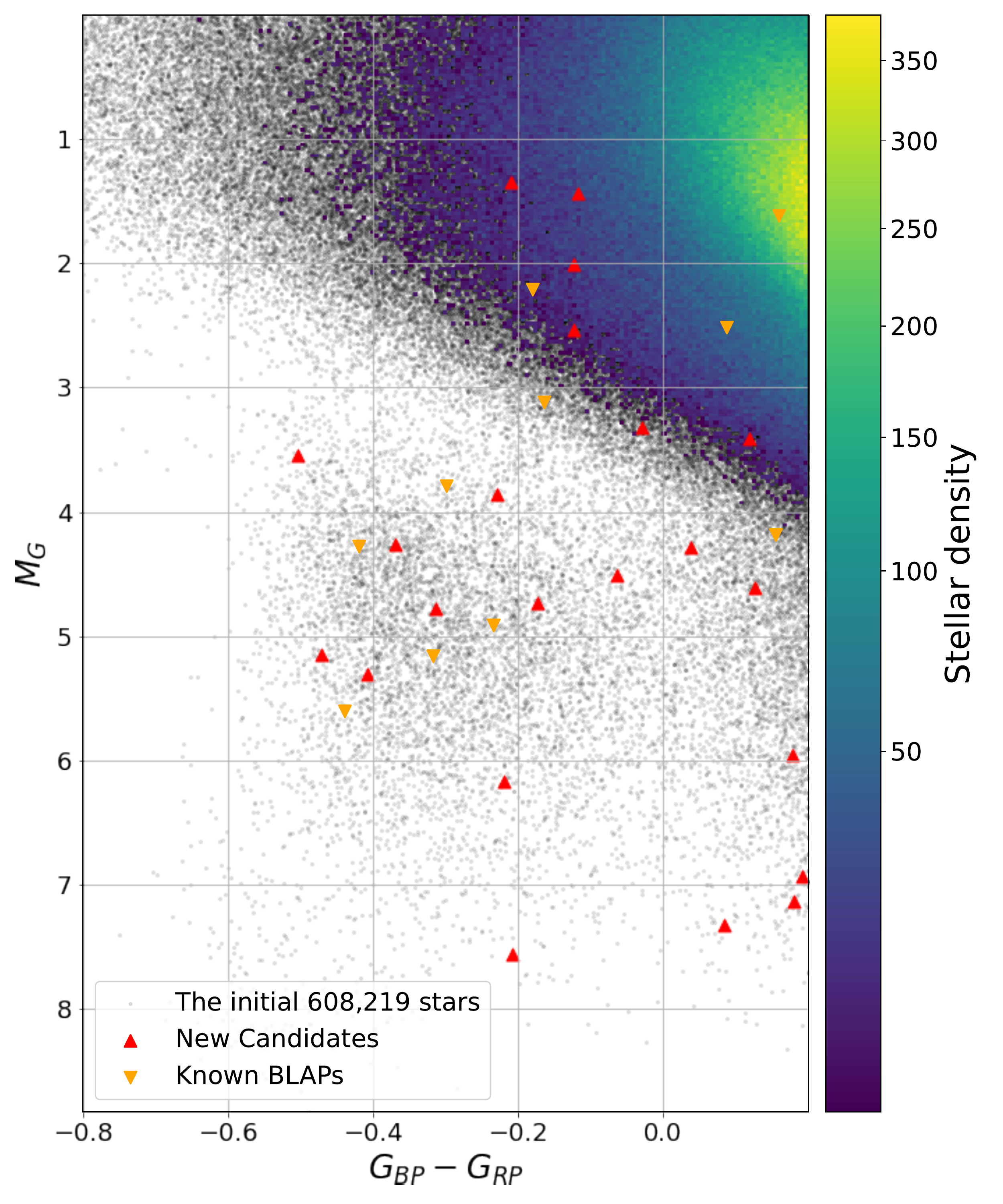}
    \caption{Hertzsprung-Russell diagram of $608,219$ stars within the selected region in the
    absolute magnitude-colour parameter space after dereddening with the method from this study.
    Due to the abundance of sources from the main sequence
    to the top right of the image, a 2D histogram is used to show the stellar density.
    Red triangles identify the $22$ new BLAP candidates in this study and the orange inverted
    triangles identify the known BLAPs with good parallaxes from GDR2 and within the PS1 footprint.
    All of them are interstellar reddening corrected.}
    \label{fig:hrdiag}
\end{figure}

The colour and absolute magnitude
selection criteria are determined from the known set of BLAPs with good parallaxes in GDR2
and within the PS1 footprint such that all of them can be
recovered. This is accomplished by applying our method to the known BLAPs to map them
onto our $\mathrm{M}_{G}$, $(G_{\mathrm{BP}}-G_{\mathrm{RP}})_0$ parameter space. We are primarily concerned
with removing contamination from the main sequence so our constraints are only applied to remove the
brighter and cooler sources from our dataset. We require $G_{\mathrm{abs, deredden}}>0.0$\,mag and
$G_{\mathrm{BP, deredden}}-G_{\mathrm{RP, deredden}}<0.2$\,mag. The constraints are loose
because of (1) the large aggregated uncertainties coming from the combined parallax
and extinction correction; (2) the infancy of our understanding of this elusive population
of blue faint pulsators. This basic colour and absolute magnitude selection
reduce the sample size to $<1\%$ of the initial size, giving us $608\,219$ subluminous and
intrinsically blue stars. These targets are then cross-matched with the ZTF DR3 to obtain
epoch photometry to determine if these targets are variable. We conservatively require our
targets to have variability $>0.2$\,mag. Sources with nearby ($50''$ if <$12$\,mag
and $10''$ if $12-15$\,mag) bright sources are removed
to minimise false positives due to contamination. As a result, only $44\,847$ are left as 
candidate subluminous, intrinsically blue variable sources. This
dereddening is also performed on the high Galactic
lattitude dataset. The reduced Galactic interstellar extinction
for the sources at these high lattitudes results in minimal contamination from main sequence
sources with $2\,704$ sources identified as candiate subluminous, intrinsically blue variable sources.
The dereddened `low parallax quality' dataset has higher uncertainties resulting 
in $31\,319$ selected sources with a higher likelihood of main sequence contamination.

The candidate variable sources may be a result of periodic or non-periodic phenomena
or contamination in the ZTF light curves. To identify light curves with
high periodicity a periodogram must be utilised. The Conditional Entropy (CE) Periodogram
has comparable computing performance to other information-based techniques but is superior
for finding periods in light curves with non-sinusoidal behaviour~\citep{2013MNRAS.434.2629G}.
It functions by identifying candidate periods where the epoch-folded data points align.
This results in a decrease in the entropy of the system as the data points are ordered. As the shape of
the alignment has no effect on the entropy, the CE periodogram is robust to a large variety of potential
signals. The algorithm computes a unit-square normalised light curve (phase and magnitude values ranging
between $0-1$). It then subdivides this space into bins and computes the quantity of data points present
in each bin for each candidate period. Highly ordered data points will fill a minimum quantity of bins
generating a peak in the CE for that frequency. Due to the binning of the data, the CE periodogram has two
additional parameters: the number of bins in the phase dimension and in the magnitude dimension.
These parameters must be optimised to determine the appropriate values for a specific dataset.
By employing a GPU-accelerated CE periodogram for the significant
improvement in processing time for large datasets~\citep{2021MNRAS.503.2665K},
these candidates are analysed
for periodicity with a period range of $3-90$\,mins, as expected from the BLAP-like
variability as reported from the theoretical front~\citep{2020MNRAS.492..232B}. 
We perform a sigma-clipping procedure on the $g_{\mathrm{ZTF}}$ and $r_{\mathrm{ZTF}}$ light curves of the candidate variable
stars with $\sigma=3$ to mask outliers from the periodogram. We utilise an oversampling factor of $5$
and subdivide the unit-square normalised epoch-folded light curves into $20$ bins in the phase
dimension and $10$ bins in the magnitude dimension as recommended by the authors of
this implementation on ZTF data.

Sources with optimal candidate periods with peak CE values greater than $10\sigma$ confidence
in the periodogram in either the $g_{\mathrm{ZTF}}$ or $r_{\mathrm{ZTF}}$ band observations
are selected. This significantly reduces the shortlist to a sample of $98$ sources
derived from both the Galactic plane dataset
and the two supplementary datasets. Manual
inspection of the epoch-folded light curves gives us the final $22$ candidate BLAPs. Six of
them are consistent with the observed periods and light curves of the known BLAPs, ten are
considered to be within the theoretical period range of BLAPs, and the last six are candidate
high-gravity BLAPs. The dereddened Hertzsprung-Russell diagram of the region selected
for this study is shown in Figure~\ref{fig:hrdiag}. 
Contamination from the main sequence dominates the top right corner of the plot with the
highest source density. The over-density centered at ($\mathrm{M}_{G}=5$,
$G_{\mathrm{BP}} - G_{\mathrm{RP}}=-0.3$) is attributed to the hot subdwarfs on
the extreme horizontal branch.
The existing BLAPs have been processed by
this method and are shown as orange inverted triangles. The 22 new candidate BLAPs are shown as
red triangles.

\begin{figure*}
    \centering
    \includegraphics[width=\textwidth]{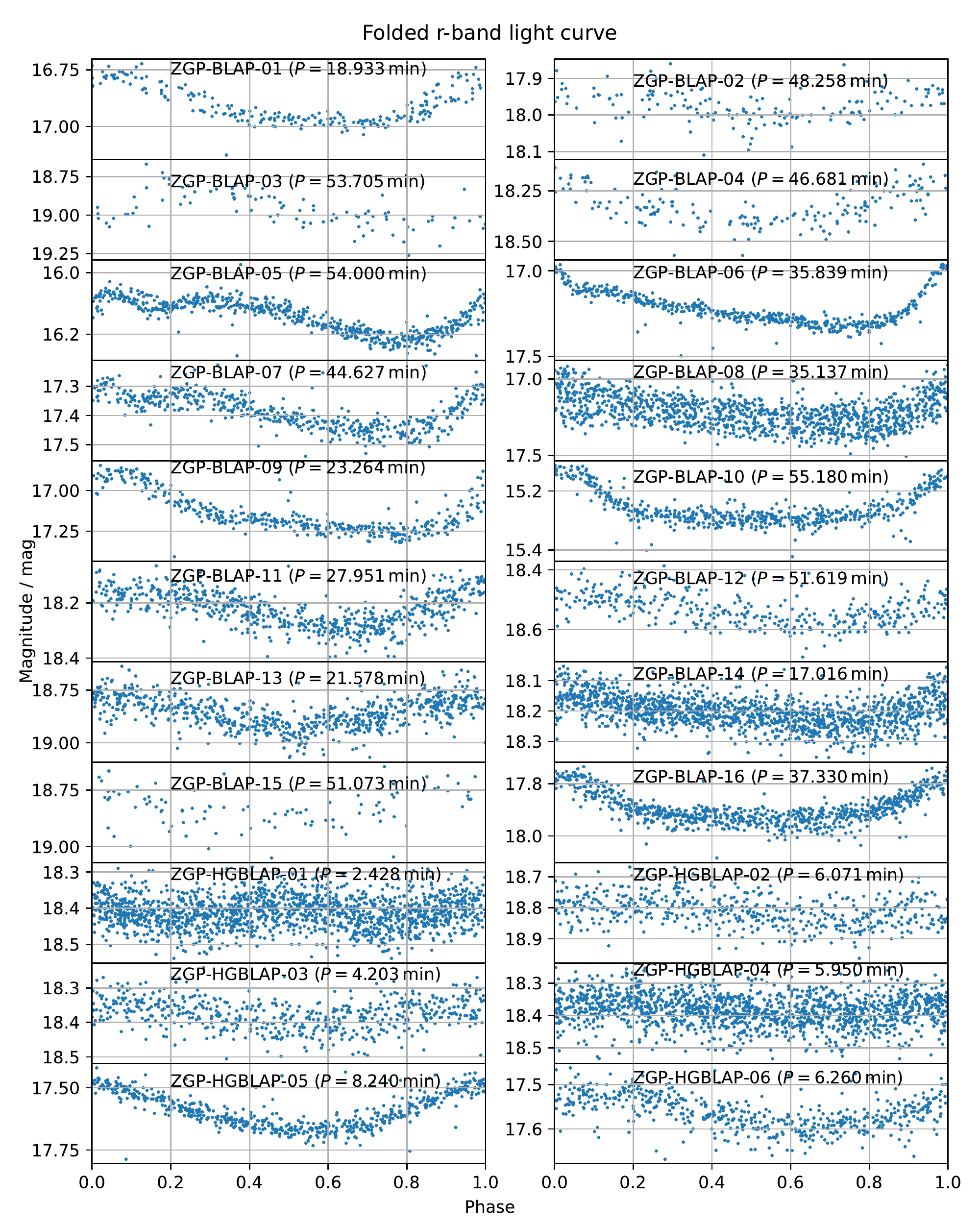}
    \caption{Epoch-folded light curves of the $22$ new candidate BLAPs ZGP-BLAP-01
    through to 16 from the top, and ZPG-HGBLAP-01 to 06 at the bottom.
    Many of the candidates show clear asymmetric
    shapes with a fast rise and slow fall consistent with the existing BLAP population as do some
    additional candidates. The High-g BLAP candidates show more symmetric light curves consistent
    with these more compact variables (see text).}
    \label{fig:ztflc}
\end{figure*}

In Figure~\ref{fig:ztflc} we display the epoch folded light curves of the 22 new candidates utilising
the periods discovered by the CE periodogram as listed in appendix A.
Most of the long period candidates show asymmetric light curves similar to those
from P17 indicative of their nature as pulsating stars. The remaining sources still demonstrate
high amplitude periodic variability which is not explained by binary companions.
The High-g BLAP candidates have symmetric light curves similar to those of K19. In this case, we
have identified these targets as candidates due to their high amplitude relative to other
hot subdwarf pulsators. The coordinates, periods and amplitudes of these stars in the $r_{\mathrm{ZTF}}$
passband used to generate this figure can be found in Appendix A.
The target we have named ZGP-BLAP-09 has been previously discovered. It was initially confused with
a nearby Red Giant Branch~(RGB) star~\citep{2020MNRAS.499.5782O}. We identified this source as
a BLAP candidate and submitted a proposal to obtain spectroscopic observations prior
to the selection of this source from GDR2 by our method.
The true classification of this source was also independently determined by observers affiliated with
the American Association of Variable Star Observers~(AAVSO) and this source was uploaded as a BLAP
candidate with the identifier ZTF J191440.84+193825.8 to the AAVSO Variable Star Index~(VSX).
Returning to the study by~\citet{2020MNRAS.499.5782O}, they also detected ZGP-BLAP-08 although
they do not make a classification.
We returned to the AAVSO VSX to determine if any of our other candidates
have been included in this catalogue.
ZGP-BLAP-11, 13 and 14 are part of Gabriel Murawski's MGAB catalogue although they do not have
firm classifications. They made a note that MGAB-V3437~(ZGP-BLAP-14)
is a `possible BLAP type variable'. None of our remaining candidate BLAPs are present on the VSX.
Finally, ZGP-HGBLAP-02 has been recently identified as a potential
single-mode sdB hot subdwarf pulsator~\citep{2021MNRAS.505.1254K}.
Due to the similarity between our High-g BLAP candidates, it is possible that the other five
candidates may also be members of this recently discovered type of pulsator.
We present follow-up spectra of $21$ of these candidate sources and propose
them as new BLAPs for future population studies.

\section{Spectroscopic Follow-up}

We carried out spectroscopic observations of ZGP-BLAP-09 using the
Optical System for Imaging and Low-Resolution Integrated
Spectroscopy~\citep[OSIRIS;][]{2000SPIE.4008..623C} in long-slit mode at the 10.4\,m
Gran Telescopio de Canarias~(GTC) on the island of La Palma, Spain.
Our program was executed on 13 September 2020 (PI: McWhirter. Program
IDs: \verb+GTC2020-159+) at an average airmass of 1.02, seeing around 0.7'',
and a waning crescent moon for 960\,s and 1200\,s with the R1000B and R2500U
gratings, respectively.

Under the proposals \verb+CL21A10+ and \verb+JL21B16+, low-resolution spectra
of the 15 BLAP candidates and 6 high gravity BLAP candidates were collected
with the SPRAT spectrograph~\citep{2014SPIE.9147E..8HP} mounted on the
Liverpool Telescope~\citep{2004SPIE.5489..679S}, also located on the island of
La Palma, Spain. The blue-optimised mode was used to collect the spectra in dark
condition with seeing at around 1'' in the summer months of 2021, see
Table~\ref{tab:blap} and \ref{tab:hgblap} in the Appendix for more details.

\begin{figure*}
    \includegraphics[width=\linewidth]{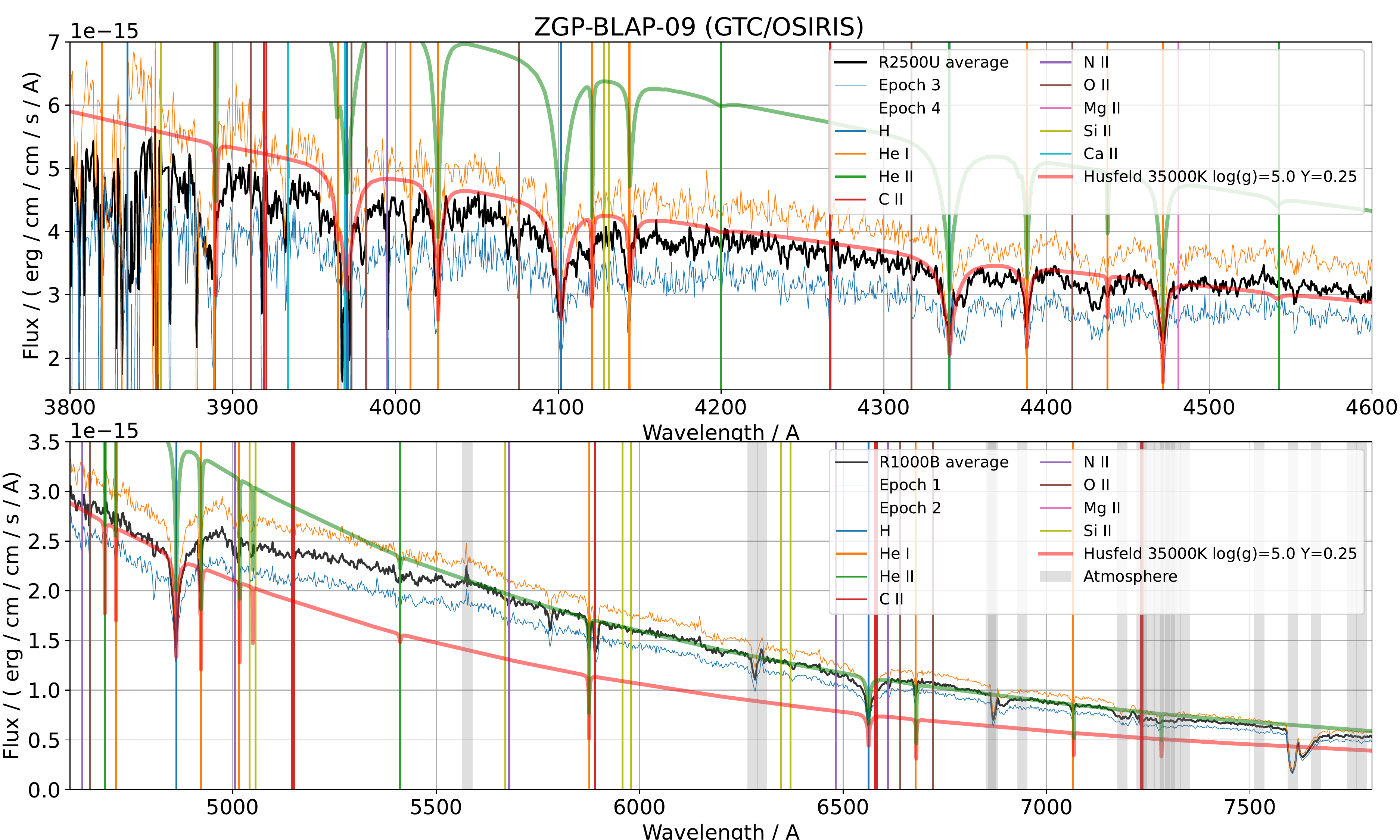}
    \caption{Top: Blue and orange spectra are GTC/OSIRIS data taken with the R2500U grating.
    The black spectrum is the weighted mean of the two. Red and green smooth spectra
    are the best fit model from \citet{1989A&A...222..150H} where the normalisations are
    best fitted with the R2500U~(red) or the R1000B~(green) spectra separately. Each fitted
    model has the same normalisation across the entire spectral range. The vertical lines
    indicate the locations of the various absorption lines from H, He, C, N, O, Mg, Si,
    Ca and residual/unsubtracted atmospheric absorption/emission lines. Note that not
    all the absorption lines are present, they are plotted because some of them are
    expected to have the same absorption intensity as the other features that clearly visible.
    Bottom: Blue and orange spectra are GTC/OSIRIS data taken with the R1000B grating.
    The black spectrum is the weighted mean of the two.}
    \label{fig:blap09}
\end{figure*}

\begin{figure*}
    \centering
    \includegraphics[width=\textwidth]{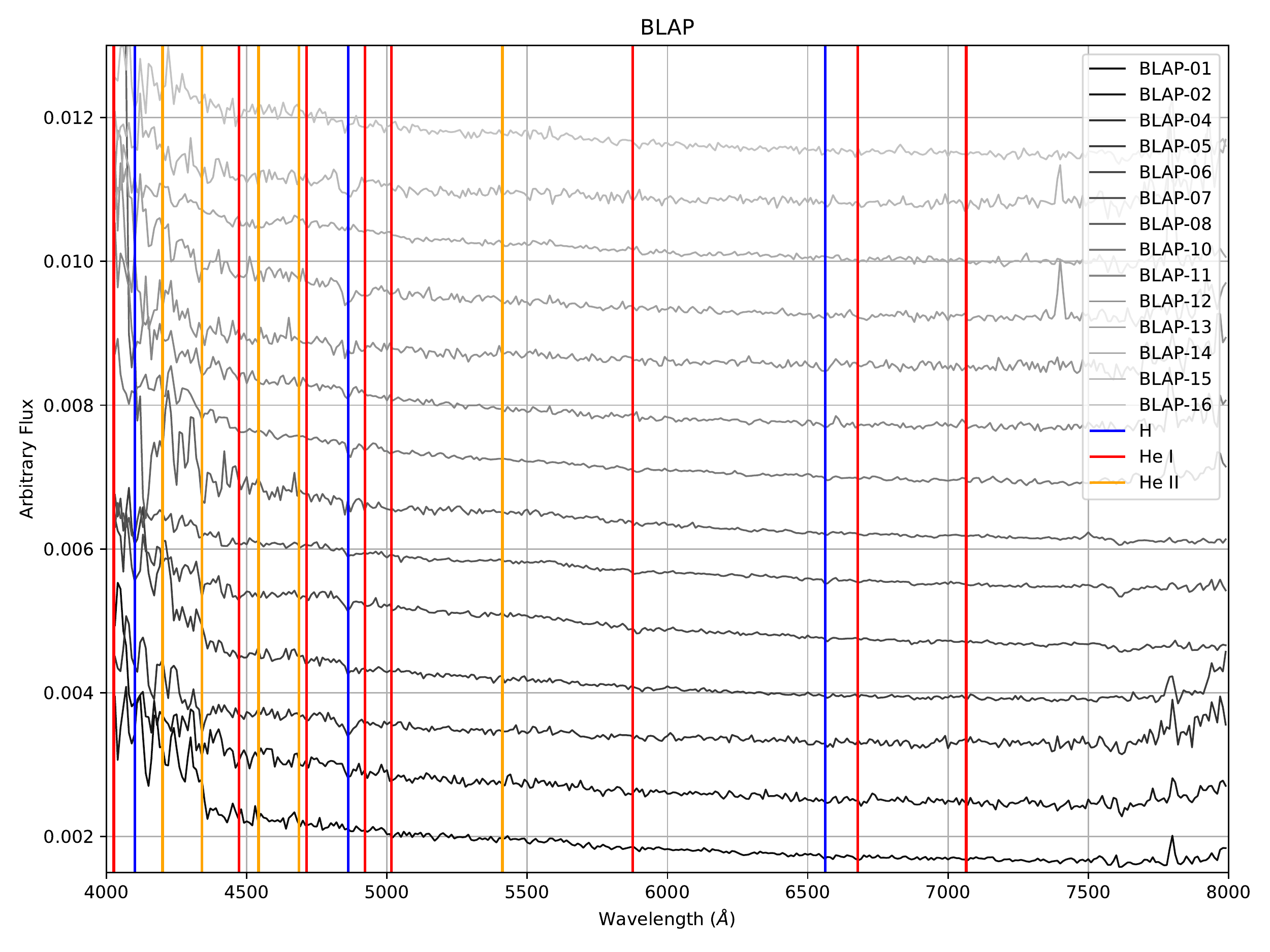}
    \caption{From bottom to top show the candidate BLAPs from 01 to 16~(except 02, 03 and 09) with a decreasing shade intensity. The spectra are the weighted average of the individual observations. The interstellar reddening is corrected the same way as the OSIRIS reduction~(Section \textsection3.1.1).}
    \label{fig:blap}
\end{figure*}

\begin{figure*}
    \centering
    \includegraphics[width=\textwidth]{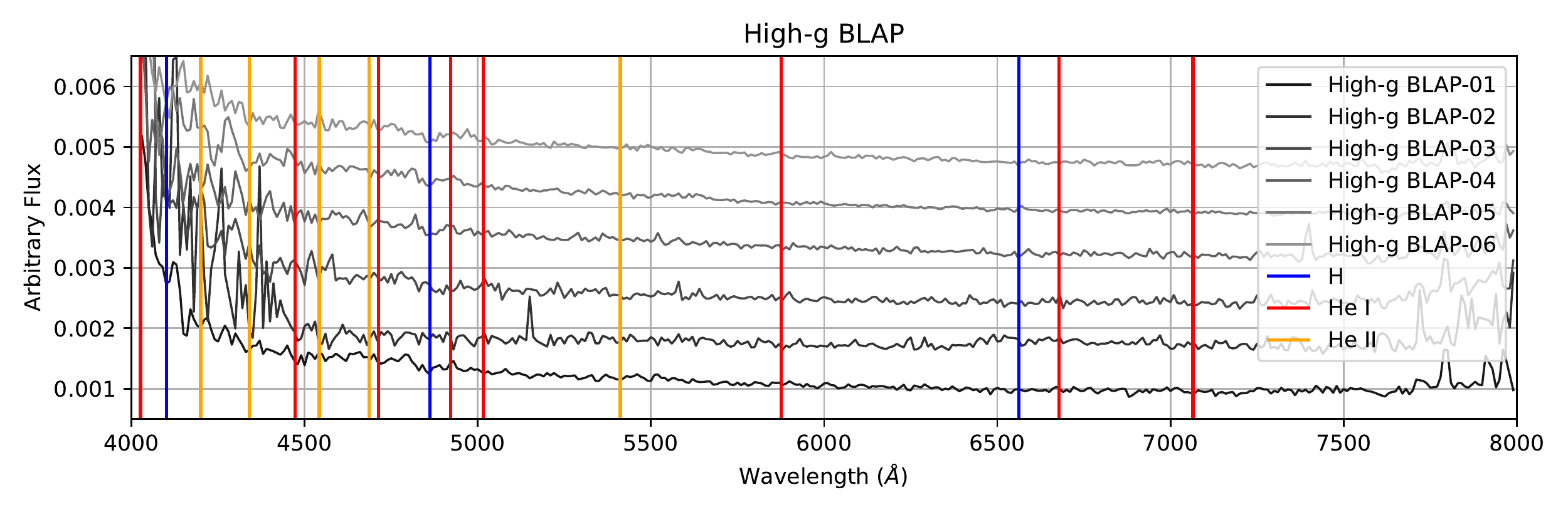}
    \caption{From bottom to top show the candidate High-g BLAPs from 01 to 06 with a decreasing shade intensity. The spectra are the weighted average of the individual observations.The spectra are the weighted average of the individual observations. The interstellar reddening is corrected the same way as the OSIRIS reduction~(Section \textsection3.1.1).}
    \label{fig:hgblap}
\end{figure*}

All the data were reduced using the new generation \textsc{iraf}-free
\textsc{ASPIRED}~\citep{2021arXiv211102127L, marco_c_lam_2021_5650127, 2020zndo...4117517V}
\footnote{\url{https://github.com/cylammarco/ASPIRED}}$^,$\footnote{\url{https://github.com/jveitchmichaelis/rascal}}
written entirely in \textsc{Python}. Standard procedure of spectral reduction was
performed: 2D spectral image reduction, tracing, optimal extraction, wavelength
and flux calibration, atmospheric absorption correction. We apply interstellar reddening correction using the values computed in Section \textsection2.1. The
reduced spectra from OSIRIS are shown in Figure~\ref{fig:blap09}, while those from
SPRAT are shown in Figure~\ref{fig:blap} and \ref{fig:hgblap}.

\subsection{BLAP candidates}
In the BLAP candidate sample, we obtained medium resolution spectra of
ZGP-BLAP-09, which we considered as the best candidate as judged from the amplitude
and period. It is the only spectrum that is corrected for the interstellar
extinction for the purpose of analysis. Low-resolution spectra were collected
with LT/SPRAT for 14 of the remaining 15 BLAP candidates. Given their relatively
large surface gravity, a grism with $R\sim$ $300$ is sufficient to resolve the
strong H and He features -- all the follow-ups show H$\beta$ absorption, most of
them also show H$\alpha$, and in a few cases the He\,I at $4472$\,\AA\ and He\,II
at $4340$\,\AA\ are present. The H$\beta$ absorption was also showing stronger feature
in the previous study using the same spectrograph~\citep{2021RNAAS...5..131M}.

\subsubsection{ZGP-BLAP-09}
The blue end is observed with the R2500U grating while the red end is with the
R1000B grating such that there is sufficient resolving power to inspect the
weaker helium lines. In fact, all the lines in the ``Strong line list'' on
NIST\footnote{\url{https://physics.nist.gov/PhysRefData/Handbook/Tables/heliumtable2.htm}}
in the range of $3800-4500$\,\AA\ can be seen. In addition, we have also identified
absorption lines from heavier elements in the spectrum, including clear absorption
features from C\,II~($4267$, $5145$, $5151$ \& $5890$\,\AA), N\,II~($3995$ \& $5680$\,\AA),
O\,II~($3911$, $3973$, $3982$, $4076$, $4317$, $4416$ \& $4649$\,\AA),
Mg\,II~($4481$\,\AA), Si\,II~($5041$ \& $6371$\,\AA) and Ca\,II~($3934$\,\AA). 
We cannot identify iron lines; or forbidden lines that are common in shell ejecta
in case the absorption comes from the environment. This is in agreement with the lack
of IR excess from the WISE mid-infrared data~\citep{2019ApJS..240...30S}.

Each individual observation does not cover the same part of a pulsation period, so
the normalisations are, as expected, different despite one is taken immediately
after another. By comparing with the medium resolution spectrum of OGLE-BLAP-001
in Figure~4 of P17, we have identified a few interesting features, from the blue
end to the red end: there are subtle hints of O\,II lines at $4076$\,\AA\ as well
as the same unidentified line at $4069$\,\AA. At $4110$\,\AA, the unidentified
absorption feature looks noise-like, but it appears on both the P17 and our data.
Towards the red wing of the H$\alpha$, there could be a C\,II line at $6578$\,\AA.
This will allow us to adopt a metallicity more appropriate for BLAPs. In the case of P17,
they adopted a metallicity typical of a hot subdwarf: solar abundances for N, S
and Fe, and one-tenth solar for C, O and Si.

Without any publicly available models that are immediately available for spectral analysis
of BLAPs, the best model grid that is suitable for preliminary fitting is the set provided
by \citet{1989A&A...222..150H}\footnote{hosted on \url{http://svo2.cab.inta-csic.es/theory/newov2/}}.
Before template fitting, we correct for the interstellar extinction based on the
PS1 dust map and the distance from \citep{2018AJ....156...58B} at $2.11$\,kpc, giving
us $E(g-r)=0.81$. We use the typical total extinction for the Galaxy, $R_V=3.1$, and
correction using the extinction law from \citet{2007ApJ...663..320F} using \citet{barbary_kyle_2016_804967}
\footnote{\url{https://github.com/kbarbary/extinction}} and $A_G = 0.789 \times A_V$~\citep{2019ApJ...877..116W}.
We find the best fit solution (from the sparsely populated parameter space) a helium rich
star~($Y=0.1$) with T$_\mathrm{eff}=35\,000$\,K and $\log(g)=5.0$ (see Figure 1).
Judging from the shape of the model~(thick red line) and the spectra, we believe at the blue
end the spectrum is already starting to plateau, a lower temperature should give a better fit.
The R1000B~(bottom) spectrum is normalised to match the flux of the R2500U~(top) spectrum at
$4600$\,\AA. By fitting the red and blue spectra separately with the normalisation as a free
parameter. If we fit with the blue spectrum and extend the model to the red, we find a
significant discrepancy in the flux between the model and the spectrum redward of the
H$\beta$ absorption line. Given the age of the model, it is possible that the model is
scaled to match the (under-subtracted) line-blanketed region of the spectrum~(the R2500U spectrum),
thus it should be interpreted as that the model agrees with the R1000B spectrum, and there
is significant flux deficit in the blue end~(thick green line). We also note that the effective
temperature of a BLAP can be strongly dependent on the phase, given the high-g BLAP reported
to have a temperature variation of $3\,000\,$K between the minimum and maximum over a period of
pulsation. Regarding the He absorption line strengths, some lines are better fitted with as
low as $Y=0.01$ model, while some lines are best fitted with the $Y=0.25$ model. We choose to
plot the only option between these two abundances, $Y=0.10$. On the other hand, the broadening
of the line shows a good agreement with the fitted surface gravity, the next step down and
up are $4.5$ and $5.5$, so among all the fitted properties, we would tentatively assign to
the \textit{average} spectrum an upper limit of T$_\mathrm{eff}<35\,000$\,K, a surface gravity
of $\log(g)=5.0\pm0.25$ and an abundant amount of helium that we are unable to constrain it to better
than $0.01<Y<0.25$.

\subsubsection{ZGP-BLAP-01, 05, 06, 14 and 16}
These five candidates have strong blue spectra, but their spectral energy distributions~(SEDs)
show mild flux deficit around $4500$\,\AA. This could be the result of the presence of
a cooler unresolved companion. This feature is the most prominent with ZGP-BLAP-14.
Should it be coming from a close companion, without considering the potential re-radiation
effect from the high energy photons from the hotter component, they are most like
late G/K-type main sequence stars. Given the absolute $G$-band magnitudes at around $5-6$\,mag,
which is about $1-2$\,mag fainter than the BLAPs, this translates to about $40-85\%$
fainter in terms of flux. Furthermore, these hypothetical main sequence companions should
dominate the flux in the red end. With the low resolving power of SPRAT, the absorption lines
could all be "washed-out" by the unresolved companions leaving these relatively smooth spectra
in the red end. With the available photometric and spectroscopic data, it is difficult to
determine their natures any further. Despite their strange SEDs, they may reveal more about
their past evolution history than the other candidates as they provide additional
features as constraints to their past and present evolution scenario.

\subsubsection{ZGP-BLAP-02, 04, 07, 08, 10, 11, 12, 13 and 15}
The remaining nine BLAP candidates show the typical blue SEDs with strong H$\beta$ absorption
and He\,II at $4340$\,\AA. The spectra are noisy in that region, but it is
statistically unlikely that all of them show the He\,II absorption, while all of the five
other candidates showing flux deficit do not show this absorption line. Using ZGP-BLAP-09 as
a proxy, we know that the $4340$\,\AA\ absorption can be very strong and comparable to the
strength of the neighbouring He-I absorption lines. These either candidates appear to be more
akin to the ``conventional'' BLAPs as were first discovered, we recommend further follow-up
with higher resolution spectra for this sample that is less likely to have a close companion
to complicate studies with additional Physics.

\subsection{High-gravity BLAP candidates}
All six High-g BLAP candidates are sowing clear H$\beta$ absorption feature. As the gravity gets
stronger, the absorption lines should be more broadened and more visible with low-resoluiton
spectra. We, however, cannot find the absorption features any more prominent than
the BLAP candidates. However, all of them are exhibiting very blue SEDs, with single-mode
pulsation period between two and nine minutes, it is difficult to assign them to any stellar
type. In the parameter space they occupy, among the existing stellar types, they can only be
high-g BLAPs, or some of the of the peculiar hot subdwarfs.
We recommend further follow-up observations with a higher resolution spectrograph.

\section{Conclusion}
We made use of the synergy of the high-level products of (1) accurate parallaxes from Gaia
DR2, (2) high cadence survey of the northern sky with the ZTF DR3, (3) the most up-to-date
3D dust map generated using Pan-STARRS 1, 2MASS and Gaia DR2, and (4) finally a
GPU-accelerated Conditional Entropy periodogram to allow the identification of the 16
BLAP candidates and 6 high gravity BLAP candidates with a desktop computer. 
Nine of the BLAP candidates
have ZTF light curves clearly exhibiting the asymmetry seen in previously observed BLAPs with periods 
consistent with current theories. We tentatively classify these candidates as BLAPs based on 
these observations. Due to the followup spectral observations, we propose to firmly
classify ZPG-BLAP-09 as a BLAP. The remaining tentative classifications are for ZPG-BLAP-05, 06, 07, 08,
10, 11 and 14. The remaining seven candidate BLAPs are consistent with the theoretical BLAP population
but could also be eccentric binary systems, $\delta$~Scuti or SX Phoenicis pulsators and therefore are too
ambiguous for a classification. All the high-gravity BLAP candidates have coherent sinusoidal 
variability as expected for these objects. ZPG-HGBLAP-02, 03, 04 and 06 have the clearest light
curves from these high-gravity candidates.
Additional spectral observations are required to determine
the nature of these variable objects as this region of the period,
amplitude and light curve symmetry feature space is currently occupied by both the high-gravity BLAPs and
a new class of radially pulsating sdB hot subdwarfs.

The best candidate, ZPG-BLAP-09, was observed with the GTC/OSIRIS in long-slit mode with grism R1000B
and R2500U. This delivered the second medium-resolution spectrum of a BLAP. The spectrum
shows moderately broadened hydrogen and helium absorption features, along with a number
of metal features that were not reported in the other medium-resolution spectrum of
a different BLAP, and certainly would not have appeared in the low-resolution spectra.
The detection of strong metal absorption features immediately suggest two scenarios,
which can be true at the same time. First, it further reinforces the case that BLAPs
have a very recent energetic past where given their large surface gravity, the heavy
elements still do not have time to set beneath the photosphere. Second, the outer layer
of the envelope is convective, and the convective cells dredge deep enough to keep
the photosphere rich in metals. Given the null detection of any emission lines so far
in all the available spectra, it is unlikely that these metals are coming from
materials being constantly accreted onto the surface.

The surface gravity and surface chemical abundances of the candidate stellar types of
BLAPs can differ by over an order of magnitude: from thermally bloated, He-enriched
stars with typical surface gravity, to the compact surface H-rich stars similar to hot
subdwarfs with moderately high surface gravity. In order to further study BLAPs, models
of their synthetic energy distribution have to be computed before we can address the
range of stellar parameters of BLAPs. As of the time of writing of this article, there
is not any publicly available models that are suitable for the analysis of BLAPs.

In this work, we believe ZGP-BLAP-09 is a genuine BLAP given the acceptable model
fitting. There are extra elements in the photosphere that were not observed before,
which can be served as a useful constraint in modelling the evolution scenario of its
progenitor as well as the mass loss process. Nine of the BLAP candidates show
promising blue spectra with clear H and He absorption lines, while other six show
flux deficit at around $4500$\,\AA. All the high-g BLAP candidates are not showing
clear absorption features, yet the spectra are indicating they are genuinely blue
objects with no evidence of companions at significantly different temperatures. We
recommend further spectral data collection of all these candidates at higher resolution,
where the eight BLAPs candidates with clear absorption features should be prioritised.

We suspect additional BLAPs await detection due to the uncertainties involved in the 
dereddening combined with limited epochs of photometry for some fields.
With the continued data releases from both ZTF and Gaia, improvements are expected for both of 
these constraints offering a potential wealth of additional discoveries in the coming years.

\section*{Acknowledgements}

PRMW acknowledges financial support from the Science and Technology Facilities Council (STFC).

PRMW thanks the useful discussions with Gavin Ramsay from the Armagh Observatory.

MCL is supported by a European Research Council (ERC) grantunder the European Union’s Horizon 2020 research and innovationprogram (grant agreement number 833031).

Based on observations obtained with the Samuel Oschin 48-inch Telescope at the Palomar
Observatory as part of the Zwicky Transient Facility project. ZTF is supported by the
National Science Foundation under Grant No. AST-1440341 and a collaboration including
Caltech, IPAC, the Weizmann Institute for Science, the Oskar Klein Center at Stockholm
University, the University of Maryland, the University of Washington, Deutsches
Elektronen-Synchrotron and Humboldt University, Los Alamos National Laboratories, the
TANGO Consortium of Taiwan, the University of Wisconsin at Milwaukee, and Lawrence
Berkeley National Laboratories. Operations are conducted by COO, IPAC, and UW.

This work has made use of data from the European Space Agency (ESA) mission
{\it Gaia} (\url{https://www.cosmos.esa.int/gaia}), processed by the {\it Gaia}
Data Processing and Analysis Consortium (DPAC,
\url{https://www.cosmos.esa.int/web/gaia/dpac/consortium}). Funding for the DPAC
has been provided by national institutions, in particular the institutions
participating in the {\it Gaia} Multilateral Agreement.

Based on observations made with the GTC telescope, in the Spanish Observatorio del
Roque de los Muchachos of the Instituto de Astrofísica de Canarias, under Director's
Discretionary Time.

The Liverpool Telescope is operated on the island of La Palma by
Liverpool John Moores University in the Spanish Observatorio del
Roque de los Muchachos of the Instituto de Astrof{\'i}sica de
Canarias with financial support from the UK Science and Technology
Facilities Council.

This research has made use of the Spanish Virtual Observatory (http://svo.cab.inta-csic.es)
supported from the Spanish MICINN/FEDER through grant AyA2017-84089

We acknowledge with thanks the variable star observations from the AAVSO International Database contributed by observers worldwide and used in this research.

\section*{Data Availability}

The data underlying this article will be shared on reasonable request to the corresponding author.




\bibliographystyle{mnras}
\bibliography{blap_zgp} 




\appendix
\section{Observation Summary}
\begin{table*}
    \centering
    \begin{tabular}{cccccccccc}
        Designation                  & RA                          & Dec                        & l                          & b                           & Period / min            & $r_{\mathrm{ZTF}}$     & $\Delta r_{\mathrm{ZTF}}$ & Instrument    & Night of Obs. \\\hline\hline
        ZGP-BLAP-01                  &  57.931979                  &  58.751071                 & 144.526097                 &   3.656876                  & 18.933                  & 16.94                  & 0.21                      & SPRAT/Blue    & 2021-08-05 \\\hline
        ZGP-BLAP-02                  &  82.905144                  &  17.536672                 & 188.005076                 &  -8.755367                  & 48.258                  & 17.98                  & 0.09                      & SPRAT/Blue    & 2021-09-02 \\\hline
        ZGP-BLAP-03                  & 169.766884                  &  13.868316                 & 239.917324                 &  64.703080                  & 53.705                  & 18.97                  & 0.26                      &      ---      &    ---     \\\hline
        ZGP-BLAP-04                  & 214.459870                  &   0.980563                 & 344.940403                 &  56.633534                  & 46.681                  & 18.33                  & 0.22                      & SPRAT/Blue    & 2021-07-03 \\\hline
        ZGP-BLAP-05$^\ast$                  & 281.940840                  & -10.435027                 &  23.214926                 &  -3.931005                  & 54.000                  & 16.13                  & 0.13                      & SPRAT/Blue    & 2021-07-01 \\\hline
        \multirow{2}{*}{ZGP-BLAP-06$^\ast$} & \multirow{2}{*}{283.187948} & \multirow{2}{*}{-5.179776} & \multirow{2}{*}{28.468139} & \multirow{2}{*}{-2.650761}  & \multirow{2}{*}{35.839} & \multirow{2}{*}{17.25} & \multirow{2}{*}{0.24}     & SPRAT/Blue    & 2021-07-01 \\
                                     &                             &                            &                            &                             &                         &                        &                           & SPRAT/Blue    & 2021-08-07 \\\hline
        \multirow{4}{*}{ZGP-BLAP-07$^\ast$} & \multirow{4}{*}{285.871177} & \multirow{4}{*}{-0.083905} & \multirow{4}{*}{34.231475} & \multirow{4}{*}{-2.718417}  & \multirow{4}{*}{44.627} & \multirow{4}{*}{17.38} & \multirow{4}{*}{0.14}     & SPRAT/Blue    & 2021-07-01 \\
                                     &                             &                            &                            &                             &                         &                        &                           & SPRAT/Blue    & 2021-08-04 \\
                                     &                             &                            &                            &                             &                         &                        &                           & SPRAT/Blue    & 2021-08-08 \\  
                                     &                             &                            &                            &                             &                         &                        &                           & SPRAT/Blue    & 2021-09-01 \\\hline         
        \multirow{2}{*}{ZGP-BLAP-08$^\ast$} & \multirow{2}{*}{288.286554} & \multirow{2}{*}{12.080933} & \multirow{2}{*}{46.138307} & \multirow{2}{*}{0.752002}   & \multirow{2}{*}{35.137} & \multirow{2}{*}{17.22} & \multirow{2}{*}{0.27}     & SPRAT/Blue    & 2021-07-01 \\
                                     &                             &                            &                            &                             &                         &                        &                           & SPRAT/Blue    & 2021-08-07 \\\hline
        \multirow{2}{*}{ZGP-BLAP-09$^\dagger$} & \multirow{2}{*}{288.670150} & \multirow{2}{*}{19.640445} & \multirow{2}{*}{53.017463} & \multirow{2}{*}{3.922213}   & \multirow{2}{*}{23.264} & \multirow{2}{*}{17.19} &  \multirow{2}{*}{0.33}    & OSIRIS/R1000B & 2020-09-13 \\
                                     &                             &                            &                            &                             &                         &                        &                           & OSIRIS/R2500U & 2020-09-13 \\\hline
        ZGP-BLAP-10$^\ast$                  & 294.099409                  &   5.084049                 &  42.654877                 &  -7.607390                  & 55.180                  & 15.28                  & 0.16                      & SPRAT/Blue    & 2021-07-03 \\\hline
        ZGP-BLAP-11$^\ast$                  & 294.693100                  & 58.698261                  &  90.781022                 & 17.144942                   & 27.951                  & 18.22                  & 0.16                      & SPRAT/Blue    & 2021-07-01 \\\hline
        ZGP-BLAP-12                  & 305.121190                  &  13.476327                 &  55.573552                 & -12.803831                  & 51.619                  & 18.52                  & 0.14                      & SPRAT/Blue    & 2021-07-03 \\\hline
        ZGP-BLAP-13                  & 312.463639                  &  33.864752                 &  76.342513                 &  -6.371979                  & 21.578                  & 18.85                  & 0.18                      & SPRAT/Blue    & 2021-07-03 \\\hline
        \multirow{2}{*}{ZGP-BLAP-14$^\ast$} & \multirow{2}{*}{317.830989} & \multirow{2}{*}{31.979516} & \multirow{2}{*}{77.847985} & \multirow{2}{*}{-11.022891} & \multirow{2}{*}{17.016} & \multirow{2}{*}{18.20} & \multirow{2}{*}{0.13}     & SPRAT/Blue    & 2021-07-02 \\
                                     &                             &                            &                            &                             &                         &                        &                           & SPRAT/Blue    & 2021-08-08 \\\hline
        ZGP-BLAP-15                  & 320.570968                  &  -0.327068                 & 52.025137	             & -33.160555                  & 51.073                  & 18.83                  & 0.18                      & SPRAT/Blue    & 2021-07-03 \\\hline
        ZGP-BLAP-16                  & 333.379362                  &  53.665594                 & 100.783160                 & -2.273963                  & 37.330                  & 17.90                  & 0.16                      & SPRAT/Blue    & 2021-07-02 \\\hline\hline
    \end{tabular}
    \caption{Observation summary of the BLAP candidates. $\dagger$ denotes the confirmed BLAP, and $\ast$ denotes the high-confidence BLAP candidates.}
    \label{tab:blap}
\end{table*}
\begin{table*}
    \centering
    \begin{tabular}{cccccccccc}
        Designation   & RA          & Dec        & l         & b         & Period / min & $r_{\mathrm{ZTF}}$ & $\Delta r_{\mathrm{ZTF}}$ & Instrument        & Night of Obs. \\\hline\hline
        ZGP-HGBLAP-01 & 282.3082215 & 41.5407333 & 71.034502 & 17.986673 & 2.428        & 18.41              & 0.12                      & SPRAT/Blue        & 2021-07-02 \\\hline
        ZGP-HGBLAP-02$^\ast$ & 286.3974935 & -6.9930426 & 28.290674 & -6.316974 & 6.071        & 18.81              & 0.12                      & SPRAT/Blue        & 2021-07-02 \\\hline
        ZGP-HGBLAP-03$^\ast$ & 288.6449108 & -7.3016160 & 29.017878 & -8.449546 & 4.203        & 18.38              & 0.11                      & SPRAT/Blue        & 2021-07-02 \\\hline
        ZGP-HGBLAP-04$^\ast$ & 295.604068  & 15.3787950 & 52.431269 & -3.920912 & 5.950        & 18.38              & 0.12                      & SPRAT/Blue        & 2021-07-03 \\\hline
        ZGP-HGBLAP-05 & 303.558858  & 41.1414719 & 78.031206 & 3.617501 & 8.240        & 17.60              & 0.18                      & SPRAT/Blue        & 2021-07-02 \\\hline
        ZGP-HGBLAP-06$^\ast$ & 327.439845  & 45.9767929 & 93.076885 & -6.099164 & 6.260        & 17.56              & 0.10                      & SPRAT/Blue        & 2021-07-02 \\\hline\hline
    \end{tabular}
    \caption{Observation summary of the High-g BLAP candidates. $\ast$ denotes the high-confidence high gravity BLAP candidates.}
    \label{tab:hgblap}
\end{table*}

\bsp	
\label{lastpage}
\end{document}